\documentclass[prd,preprint,showpacs]{revtex4}
\usepackage{amsmath}

\newcommand{\be}{\begin{equation}}
\newcommand{\ee}{\end{equation}}
\newcommand{\bea}{\begin{eqnarray}}
\newcommand{\eea}{\end{eqnarray}}

\begin{document}

\title{Statefinders, higher-order energy conditions, and sudden future singularities.}
\author{Mariusz P. D\c{a}browski}
\email{mpdabfz@sus.univ.szczecin.pl}
\affiliation{Institute of Physics, University of Szczecin,
Wielkopolska 15, 70-451 Szczecin, Poland.}

\date{\today}

\begin{abstract}
We link observational parameters such as the deceleration parameter, the jerk, the kerk
(snap) and higher-order derivatives of the scale factor, called statefinders, to the
conditions which allow to develop sudden future singularities of pressure with finite
energy density. In this context, and within the framework of Friedmann cosmology,
we also propose higher-order energy conditions which relate time derivatives of the
energy density and pressure which may be useful in general relativity.
\end{abstract}


\maketitle

The current observational data from type Ia supernovae
\cite{supernovae} favours the cosmological models of dark energy
of the phantom type \cite{phantom}. Phantom is the matter which
has a very strong negative pressure which violates the null energy
condition $\varrho + p >0$ ($p$ - the pressure, $\varrho$ - the energy density)
and, consequently, all the remaining energy conditions \cite{he}.
It has very severe physical consequences since it
leads to classical and quantum instabilities \cite{instab}. It is
even argued that it should not be accepted and that there should
be some other explanation of the phenomenon of dimming of
supernovae \cite{kaloper}. On the other hand, the observational
data is interpreted only in terms of the simple framework of
Friedmann cosmology with an additional restriction of the
barotropic form of the equation of state $p = w \varrho$ ($w=$ const.) which
links the pressure and the energy density of matter. However, in
general, the equation of state does not necessarily have to be of
barotropic type. The most interesting possibility is that it changes
with time during the evolution of the universe which makes the
barotropic index $w = w(t)$ to be the function of time \cite{w(t),onemli}.

From the observational point of view one is able to expand the equation of state
in series around the present moment of the evolution by taking appropriate
derivatives of the pressure with respect to the energy density.
This procedure, however, requires the usage of the time
derivatives of the scale factor up to suitable orders which would
show the "tendency" of the universe to change its equation of
state in future or study its evolution in the past. Due to this,
the observational parameters made of the time
derivatives of the scale factor were named {\it statefinders} \cite{jerk}.
Of course, in a general case, the spatial derivatives of all the
kinematic quantities in non-isotropic models may also be involved,
but in this paper we restrict ourselves to isotropic models only.
In fact, statefinders generalize such well-known observational
characteristics of the expansion as the Hubble (first-order) and the
deceleration (second-order) parameters. The third-order
characteristics was called jerk \cite{jerk} while the fourth-order
characteristic was called either snap \cite{snap} or kerk
\cite{genphan}. In this paper we discuss even higher-order
characteristics with the suggested names "lerk", "merk" etc. with
the logical numbering them alphabetically with an addition of the
term "erk" \cite{genphan}.

Phantom models of the universe ($w<-1$) admit a new type of singularity
which is called Big-Rip (BR) \cite{phantom}. At Big-Rip the energy density and
pressure diverge as a result of having the infinite value of the scale factor $a(t)$
at finite time. This is different from the ordinary Big-Crunch (BC) singularity
which leads to a blow-up of the energy density and pressure for
the scale factor approaching zero value at finite time. Due to
time symmetry the Big-Rip time singularity could have also
appeared in the past but it is of course different from ordinary Big-Bang
(BB) singularity.

However, if one drops the assumption of an exact equation of state
in the field equations one is able to face another type of singularity in the
universe, which is called a sudden future (SF) singularity
\cite{barrow04,barrow041,stefancic,nojiri}.
This is a singularity of pressure only, with finite energy density of matter.
Unlike phantom, it appears for the matter fulfilling the strong
energy condition $\varrho > 0, \varrho + 3p > 0$, the weak energy condition
$\varrho > 0, \varrho + p > 0$, though violating the
dominant energy condition $\varrho >0,  -\varrho < p < \varrho$ \cite{lake,barrow042,sfs1}.
The nature of SF singularities in terms of geodesic completeness has also been
discussed \cite{lazkoz04}.

In this paper we discuss SF singularities \cite{barrow04,barrow041}
and generalized SF singularities \cite{barrow042} and relate their
emergence to observational parameters composed of higher-order
derivatives of the scale factor. These, on the other hand, may
give some insight into the problem of the (changing in time) equation of state of the
matter in the universe, since they are directly related to higher-order time derivatives
of the energy density and pressure according to the expansion
\bea
\label{series}
p &=& p_0 + \frac{dp}{d\varrho}\mid_0 (\varrho - \varrho_0) + \frac{1}{2!}
\frac{d^2p}{d\varrho^2}\mid_0 (\varrho - \varrho_0)^2 + O\left[(\varrho -
\varrho_0 )^3\right] \nonumber \\
&=&
p_0 + \frac{\dot{p}_0}{\dot{\varrho}_0} (\varrho - \varrho_0)
+ \frac{1}{2!}
\frac{\ddot{p}_0\dot{\varrho}_0 - \dot{p}_0\ddot{\varrho}_0}{\dot{\varrho}^3}
(\varrho - \varrho_0)^2 + O\left[(\varrho - \varrho_0 )^3\right]~,
\eea
where index "0" refers to a quantity taken at the current moment
of the evolution $t=t_0$.

The SF singularities appear in the simple framework of
Einstein-Friedmann cosmology governed by the standard field
equations (we have assumed that $8\pi G = c =1$)
\bea
\label{rho}
\varrho &=& 3 \left(\frac{\dot{a}^2}{a^2} + \frac{K}{a^2}
\right)~,\\
\label{p}
p &=& - \left(2 \frac{\ddot{a}}{a} + \frac{\dot{a}^2}{a^2} + \frac{K}{a^2}
\right)~.
\eea
together with the energy-momentum conservation law
\bea
\label{dotrho}
\dot{\varrho} &=& - 3 \frac{\dot{a}}{a} \left(\varrho + p \right)~,
\eea
$a(t)$ is the scale factor, $K =0, \pm 1$ is the curvature index,
$\varrho$ is the energy density, and $p$ is the pressure.

From (\ref{p}) one can see \cite{barrow04} that the singularity of
pressure $p \to \infty$ occurs when acceleration $\ddot{a} \to -
\infty$, while $a$ and $\dot{a}$ are regular. This can be achieved for the scale factor
of the form
\bea
\label{sf2}
a(t) &=& A + \left(a_s - A \right) \left(\frac{t}{t_s}\right)^m -
A \left( 1 - \frac{t}{t_s} \right)^n~,
\eea
where $a_s \equiv a(t_s)$ with $t_s$ being the SF singularity time and $A, m, n =$ const.
It is obvious from (\ref{sf2}) that $a(0)=0$ and so at zero of time
a BB singularity develops. For the sake
of further considerations it is useful to write down the higher-order derivatives
of the scale factor (\ref{sf2}), i.e.,
\bea
\label{dota}
\dot{a} &=& \frac{m}{t_s} \left(a_s - A\right)
\left(\frac{t}{t_s}\right)^{m-1} + A \frac{n}{t_s}\left( 1 - \frac{t}{t_s}
\right)^{n-1}~,\\
\label{ddota}
\ddot{a} &=& \frac{m\left(m-1\right)}{t_s^2}\left(a_s -A \right) \left(\frac{t}{t_s}\right)^{m-2}
\nonumber \\
&-& A \frac{n(n-1)}{t_s^2}\left( 1 - \frac{t}{t_s} \right)^{n-2}~,\\
\label{dddota}
\dddot{a} &=& \frac{m\left(m-1\right)\left(m-2\right)}{t_s^3}\left(a_s -A \right)
\left(\frac{t}{t_s}\right)^{m-3} \nonumber \\
&+& A \frac{n(n-1)(n-2)}{t_s^3}\left( 1 - \frac{t}{t_s} \right)^{n-3}~,
\eea
or, for a general time derivative of an order $r$:
\bea
\label{dotageneral}
a^{(r)} &=& \frac{m(m-1)(m-2)...(m-r+1)}{t_s^r} \left(a_s -A \right) \left(\frac{t}{t_s}
\right)^{m-r} \nonumber \\
&+& (-1)^{r-1} A \frac{n(n-1)(n-2)...(n-r+1)}{t_s^r}\left( 1 - \frac{t}{t_s} \right)^{n-r}~
\eea

The main point is that the evolution of the Universe, as described
by the scale factor (\ref{sf2}), begins with the standard BB
singularity at $t=0$ and terminates at SF singularity for $t=t_s$
provided we choose
\bea
\label{nq}
1 < n < 2, &\hspace{0.3cm}& 0<m \leq 1~.
\eea
For these values of $n$ and $m$ the scale factor (\ref{sf2})
vanishes and its derivatives (\ref{dota})-(\ref{ddota}) diverge
at $t=0$ leading to the divergence of $\varrho$ and $p$ in
(\ref{rho})-(\ref{p}) (BB singularity). On the other hand, the scale factor (\ref{sf2})
and its first derivative (\ref{dota}) remain constant while its
second derivative (\ref{ddota}) diverge leading to a divergence of
pressure in (\ref{p}) only with {\it finite energy density}
(\ref{rho}). It has been shown \cite{barrow042} that more general
sudden future singularities appear, provided we choose
\bea
\label{Nnq}
N < n < N+1, &\hspace{0.3cm}& 0<m \leq 1~,
\eea
instead of (\ref{nq}). It means that for any integer $N+1 = r$ we have a
singularity in the scale factor derivative $a^{(r)}$ (cf.
(\ref{dotageneral})), and consequently in the appropriate pressure derivative $p^{(r-2)}$.
This, for any $r \geq 3$, gives a sudden future singularity which
{\it fulfills all the energy conditions} including the dominant
one \cite{lake}.

Now, we introduce the higher-order characteristics or statefinders
\cite{jerk,snap,genphan} and relate
them to the emergence of sudden future singularities.
We also formulate the higher-order energy conditions which may be
useful in general relativity.

The well-known characteristics of the universe expansion are:
the Hubble parameter
\bea
\label{hubb}
H &=& \frac{\dot{a}}{a}~,
\eea
and the deceleration parameter
\bea
\label{dec}
q  &=&  - \frac{1}{H^2} \frac{\ddot{a}}{a} = - \frac{\ddot{a}a}{\dot{a}^2}~,
\eea
while the new characteristics are: the jerk parameter \cite{jerk}
\bea
\label{jerk}
j &=& \frac{1}{H^3} \frac{\dddot{a}}{a} =
\frac{\dddot{a}a^2}{\dot{a}^3}~,
\eea
and the "kerk" (snap) parameter \cite{snap,genphan}
\bea
\label{kerk}
k &=& -\frac{1}{H^4} \frac{\ddddot{a}}{a} =
-\frac{\ddddot{a}a^3}{\dot{a}^4}~,
\eea
the "lerk" parameter
\bea
\label{lerk}
l &=& \frac{1}{H^5} \frac{a^{(5)}}{a} =
\frac{a^{(5)} a^4}{\dot{a}^5}~,
\eea
and "merk", "nerk", "oerk", "perk" etc. parameters, of which a general term
may be expressed as
\bea
\label{dergen}
x^{(i)} &=& (-1)^{i+1}\frac{1}{H^{i}} \frac{a^{(i)}}{a} = (-1)^{i+1}
\frac{a^{(i)} a^{i-1}}{\dot{a}^{i+1}}~,
\eea
and its time derivative reads as
\bea
\label{dergentd}
\left(x^{(i)}\right)^{\cdot} &=& H \left[i(q+1)x^{(i)} - \left( x^{(i+1)} +
x^{(i)} \right) \right]~.
\eea

A possible {\it blow-up of statefinders} may easily be linked to an
emergence of singularities. In particular, this may be the
signals for SF singularities. In fact, it is possible to formulate
higher-order energy conditions which may be related to
statefinders. We will study this in what follows.

The application of the definitions of the parameters (\ref{hubb}), (\ref{dec}),
{\ref{jerk}), (\ref{kerk}), and (\ref{lerk}) gives the following equalities
for the time derivatives of the Hubble parameter \cite{genphan}:
\bea
\label{dotH}
\dot{H} &=& -H^2 \left( q+1 \right)~,\\
\label{ddotH}
\ddot{H} &=& H^3 \left( j+3q+2 \right)~,\\
\label{dddotH}
\dddot{H} &=& -H^4 \left[ k+4j+3q \left( q+4 \right)+6 \right]~,\\
\label{ddddotH}
\ddddot{H} &=& H^5 \left[l+5k+10j(q+2)+30q(q+2)+24 \right]~.
\eea
Using the equations (\ref{rho}), (\ref{p}), (\ref{dotrho}), (\ref{hubb}) and (\ref{dec})
the three energy conditions (weak, strong, and dominant) are equivalent to
\bea
\label{weak0}
\varrho + p & = & 2 H^2 \left(q + 1 + \frac{K}{a^2H^2} \right)>0,\hspace{0.5cm} \varrho>0~,\\
\label{dom0}
\varrho - p & = & 2 H^2 \left(-q + 2 + 2\frac{K}{a^2H^2} \right)>0,\hspace{0.5cm} \varrho>0~,\\
\label{strong0}
\varrho + 3p &=& 6qH^2 > 0,\hspace{0.5cm} \varrho>0~.
\eea
In fact, the weak energy condition requires both (\ref{weak0}) and (\ref{dom0}) to hold.
Notice that the application of (\ref{rho}), (\ref{p}) and (\ref{dotrho})
gives
\bea
\label{dotHrp}
\dot{H} &=& \frac{K}{a^2} - \frac{1}{2} \left( \varrho + p \right)~,
\eea
which, after further differentiation with respect to time, and the
usage of the formulas (\ref{dotH})-(\ref{ddddotH}), may give the
higher-order {\it weak} energy conditions which relate time derivatives
of the energy density $\varrho$ and the pressure $p$. These are
\bea
\label{weak1st}
\dot{\varrho} + \dot{p} &=& - 2H^3 \left(j + 3q + 2 + 2\frac{K}{a^2H^2} \right)>0~,
\hspace{0.5cm} \dot{\varrho}>0~\\
\label{weak2nd}
\ddot{\varrho} + \ddot{p} &=& 2H^4 \left[k + 4j + 3q(q+4) + 6
\right. \nonumber \\
&& \left. + 2\frac{K}{a^2H^2}\left(q + 3 \right) \right]>0~,\hspace{0.5cm} \ddot{\varrho}>0~\\
\label{weak3rd}
\dddot{\varrho} + \dddot{p} &=& - 2H^5 \left[l + 5k + 10 j (q+2) + 30 q(q+2) + 24 \right.
\nonumber \\
&& \left. + 2 \frac{K}{a^2H^2} \left(j + 9q + 12 \right) \right]>0,
\hspace{0.5cm} \dddot{\varrho}>0~~.
\eea
The application of (\ref{weak0}) and
(\ref{weak1st}) allows to write the time derivatives of
pressure
\bea
\label{dotp}
\dot{p} &=& 2H^3 \left( - j + 1 + \frac{K}{a^2H^2} \right)~\\
\label{ddotp}
\ddot{p} &=& 2H^4 \left[k + j - 3(q+1) -
\frac{K}{a^2H^2}(q+3)\right]~,\\
\label{dddotp}
\dddot{p} &=& -2H^5 \left[l + 2k + j(q-1) - 6q^2 - 21q - 12 \right. \nonumber \\
&& \left. - \left( j + 9q + 12 \right)\frac{K}{a^2H^2} \right]~.
\eea
From (\ref{dotp}), (\ref{ddotp}) and (\ref{dddotp})
one can see that the consecutive SF singularities as given by the
condition (\ref{Nnq}) are {\it strictly related} to a blow-up of
the statefinders. In fact, for $N=1$ there is blow-up of the deceleration parameter $q$
($p$ diverges), for $N=2$ there is a blow-up of the jerk $j$ ($\dot{p}$ diverges) etc.

Now we can write down the higher-order {\it dominant and strong} energy
conditions, i.e,
\bea
\label{dom1st}
\dot{\varrho} - \dot{p} &=& -2H^3 \left( - j + 3q + 4 + 4
\frac{K}{a^2H^2} \right)>0~,\hspace{0.5cm} \dot{\varrho}>0\\
\label{strong1st}
\dot{\varrho} + 3 \dot{p} &=& - 6 H^3 (q+j)>0~,\hspace{0.5cm} \dot{\varrho}>0\\
\label{dom2nd}
\ddot{\varrho} - \ddot{p} &=& 2H^4 \left[-k + 2j + 3q(q+6) + 12 \right. \nonumber
\\ && \left. + 4 \frac{K}{a^2H^2}(q+3)\right]>0~,\hspace{0.5cm} \ddot{\varrho}>0\\
\label{strong2nd}
\ddot{\varrho} + 3 \ddot{p} &=& 6H^4 \left[k + 2j + q(q+2)
\right]>0~,\hspace{0.5cm} \ddot{\varrho}>0 \\
\label{dom3rd}
\dddot{\varrho} - \dddot{p} &=& - 2H^5 \left[-l + k + 2j (4q + 11) + 6(7q^2 + 17q + 8)
\right. \nonumber \\
&& \left. + 4 \left( j + 9q + 12 \right)\frac{K}{a^2H^2}
\right]>0~, \hspace{0.5cm} \dddot{\varrho}>0\\
\label{strong3rd}
\dddot{\varrho} + 3 \dddot{p} &=& - 6H^5 \left[l + 3k + 2j (5q +
3) + 6q (q+1) \right]>0, \hspace{0.5cm} \dddot{\varrho}>0~.
\eea

Referring to SF singularities, one can see that it is not possible
to fulfil any generalized dominant energy condition if any of
statefinders $q, j, k, l$ etc. is singular. This is because in the
appropriate expressions (\ref{weak0}) and (\ref{dom0}),
(\ref{weak1st}) and (\ref{dom1st}) etc. the signs in front of
corresponding statefinders are the opposite.

Having given the higher-order time derivatives of the
energy density and pressure related to statefinders, one is able
to invent some {\it more sophisticated} energy conditions which may be
obeyed by presumably singular cosmological models
\cite{barrow042,lake}. As an example consider an energy condition
\bea
\alpha \varrho &>& \dot{p}~,
\eea
with $\alpha=$ const., which, after the application of (\ref{rho})
and (\ref{dotp}) gives
\bea
j &>& \left(1 - \frac{3\alpha}{2H} \right)\left(1 + \frac{K}{a^2H^2}
\right)~,
\eea
and this may prevent the emergence of SF singularity for $N=2$ in
(\ref{Nnq}).

Finally, the observational determination of the value of jerk by
using type Ia supernovae sample has been performed
\cite{riess2004} and it claims that $j_0 > 0$. Similar
investigations may perhaps also be possible for higher-order
characteristics such as kerk/snap, lerk etc. Some hints about
that are given in \cite{snap}.

\begin{center}
{\bf Acknowledgments.}
\end{center}

I appreciate comments from T. Chiba, S. Nojiri, V.K. Onemli, V. Sahni and H. \v{S}tefan\v{c}i\'{c}.

\end{document}